\documentstyle[twocolumn,ifthen,floats,aps,prl,epsfig]{revtex}

\begin{document}
\draft

\newboolean{@epsimage}
\setboolean{@epsimage}{true}

\twocolumn[\hsize\textwidth\columnwidth\hsize\csname@twocolumnfalse\endcsname%

\title{Coherent electron transport in a Si quantum dot dimer}

\author{L.P. Rokhinson, L.J. Guo\cite{guoaddr}, S.Y. Chou, and D.C. Tsui}
\address{Department of Electrical Engineering, Princeton University,
Princeton, NJ 08544, USA}

\author{E. Eisenberg$^{\dag\ddag}$, R. Berkovits$^{\dag\ddag\S}$ and B.L.
Altshuler$^{\dag\ddag}$}
\address{$^{\dag}$Department of Physics, Princeton
University, Princeton, NJ 08544, USA\\$^{\ddag}$NEC Research
Institute, 4 Independence Way, Princeton, NJ 08540,
USA\\$^{\S}$Minerva Center and Department of Physics, Bar-Ilan
University, Ramat-Gan 52900, Israel}

\date{\today}

\maketitle

\begin{abstract}
We show that the coherence of charge transfer through a weakly
coupled double-dot dimer can be determined by analyzing the
statistics of the conductance pattern, and does not require large
phase coherence length in the host material. We present an
experimental study of the charge transport through a small Si
nanostructure, which contains two quantum dots. The transport
through the dimer is shown to be coherent. At the same time, one
of the dots is strongly coupled to the leads, and the overall
transport is dominated by inelastic co-tunneling processes.
\end{abstract}


\vskip1pc]

The ability to preserve quantum coherence over large distances and
during extended period of time plays a key role in the quest for
alternative schemes for conventional electronics and nonclassical
electronic behavior. For this purpose, low-dimensional structures,
in particular quantum dots, have an obvious advantage, because the
$k$-space for inelastic scattering events is reduced due to the
reduced dimensionality. Conventionally, coherence is probed by
quantum interference effects, such as weak localization. In closed
quantum dots, coherence has been investigated by embedding the dot
in one arm of an Aharonov-Bohm (AB) interferometer\cite{yacoby95}.
This method requires a host material with large phase coherence
length $l_{\phi}$, larger than the total length of both arms of
the interferometer.

Is it possible to measure coherence in a mesoscopic device
embedded in a material with small $l_{\phi}$? We encountered this
problem during our studies of Si nanostructures, where in the host
two-dimensional electron gas $l_{\phi}<1000$ \AA\ and
interferometric methods cannot be used. Coherence in Si
nanostructures is of particular interest because Si has
intrinsically long spin relaxation time, which is of great
importance for future spintronic devices. In this work we present
a new method to discriminate between coherent and incoherent
transport through a double dot system. The method is based on
statistical analysis of the conductance pattern, and does not rely
on large $l_{\phi}$ in the surrounding contact regions.

Statistical properties of single quantum dots have been
extensively investigated over the past ten
years\cite{qd-stat-review}. In a weakly coupled double-dot dimer
the statistical properties of each dot, such as the peak heights
distribution, are almost uncorrelated. However, the way the
individual conductances are combined into the total conductance of
the dimer depends on whether the transport is coherent or
sequential.  Thus, a proper deconvolution of the total conductance
can identify the type of the transport through the whole
structure. This method directly probes the coherence  during the
charge transfer through several nanostructures, which is of a
paramount importance for any practical applications.

In the following, we analyze charge transport through a Si
double-dot device. We show that electrons are transferred
coherently through the dimer, even though the transport is
dominated by inelastic co-tunneling processes. This result was not
anticipated {\it a priori}, since conventional wisdom associates
inelastic processes with decoherence.

\begin{figure}[tb]
\def\ffile{cond}
\ifthenelse{\boolean{@epsimage}}
{\ifthenelse{\boolean{@twocolumn}}
{\epsfig{file=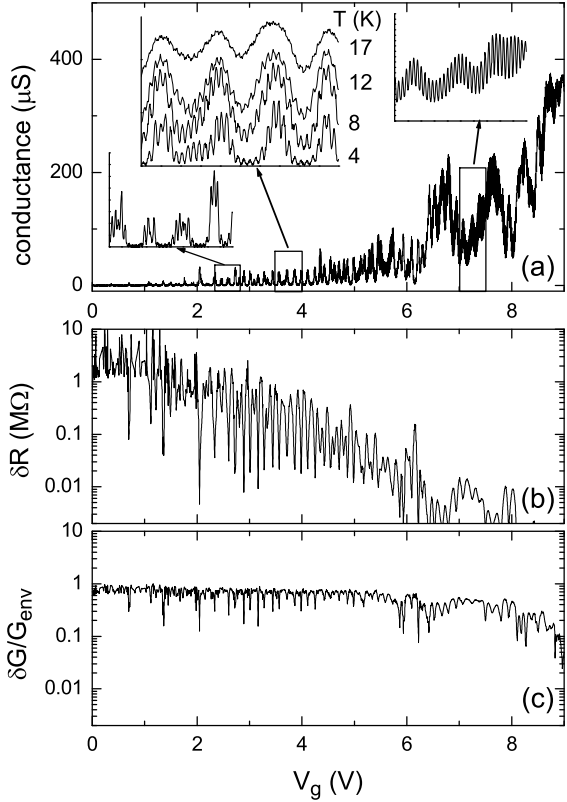,width=3.25in}}
{\centering\epsfig{file=\ffile.eps,width=5in}}}
{\ifthenelse{\boolean{@twocolumn}}
{\special{isoscale \ffile.wmf, 3.5in 5.0in}\vspace{4.8in}}
{\special{isoscale \ffile.wmf, 5.25in 6.5in}\vspace{6.5in}}}
\caption{a) Conductance as a function of gate voltage measured at
$T=1.8$ K. Some regions are enlarged in the insets. In b) and c)
the amplitude of the fast oscillations is extracted from the curve
in a) in units of resistance $\delta R$ and in units of
conductance, normalized by the envelope of the total conductance,
$\delta G/G_{\rm env}$.}
\label{\ffile}
\end{figure}

\begin{figure}[tb]
\def\ffile{trans}
\ifthenelse{\boolean{@epsimage}}
{\ifthenelse{\boolean{@twocolumn}}
{\epsfig{file=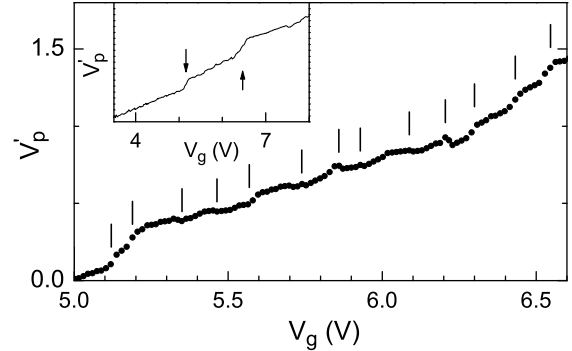,width=3.25in}}
{\centering\epsfig{file=\ffile.eps,width=5in}}}
{\ifthenelse{\boolean{@twocolumn}}
{\special{isoscale \ffile.wmf, 3.5in 2.4in}\vspace{2.4in}}
{\special{isoscale \ffile.wmf, 5.25in 4.5in}\vspace{4.5in}}}
\caption{Normalized peak position of fast oscillations,
$V_p^{'}=[V^p_g(N)/\langle\Delta V_{g2}\rangle-N]$, where
$V_g^p(N)$ is the position of the $N$-th peak and $\langle\Delta
V_{g2}\rangle=14$ mV, is plotted as a function of $V_g^p(N)$. $N$
is chosen to set $V_p^{'}=0$ at $V_g=5$ V. Vertical lines mark CB
peaks positions in dot 1. In the inset the phase is plotted for a
wider range of $V_g$ and two large slips, attributed to the
charging of traps, are marked with arrows.}
\label{\ffile}
\end{figure}

\begin{figure}[tb]
\def\ffile{histo}
\ifthenelse{\boolean{@epsimage}}
{\ifthenelse{\boolean{@twocolumn}}
{\epsfig{file=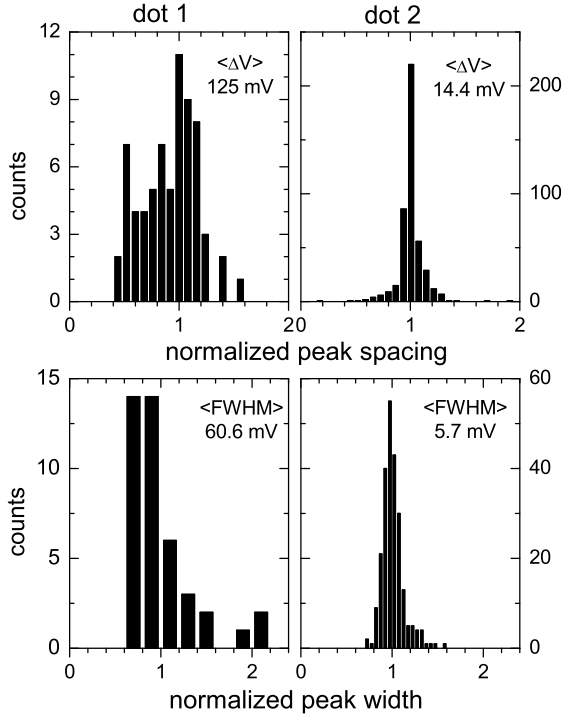,width=3.25in}}
{\centering\epsfig{file=\ffile.eps,width=5in}}}
{\ifthenelse{\boolean{@twocolumn}}
{\special{isoscale \ffile.wmf, 3.5in 4.3in}\vspace{4.3in}}
{\special{isoscale \ffile.wmf, 5.25in 6.5in}\vspace{6.5in}}}
\caption{Histograms of peak spacing (top) and width (bottom) are
plotted for dot 1 (left) and dot 2 (right).}
\label{\ffile}
\end{figure}

\begin{figure}[tb]
\def\ffile{scalings}
\ifthenelse{\boolean{@epsimage}}
{\ifthenelse{\boolean{@twocolumn}}
{\epsfig{file=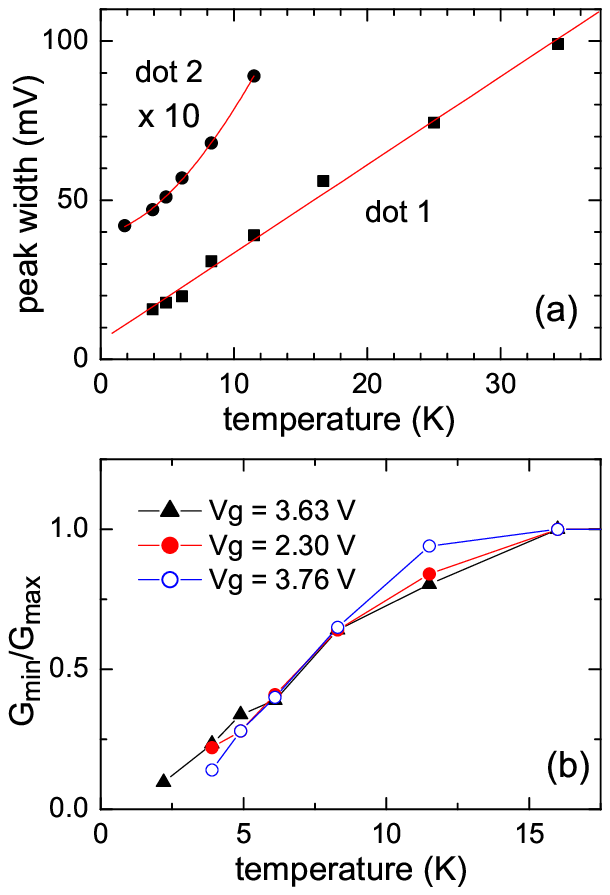,width=3.25in}}
{\centering\epsfig{file=\ffile.eps,width=5in}}}
{\ifthenelse{\boolean{@twocolumn}}
{\special{isoscale \ffile.wmf, 3.25in 4.in}\vspace{4.in}}
{\special{isoscale \ffile.wmf, 5.25in 6.5in}\vspace{6.5in}}}
\caption{a) Scaling of the peak full width at half maximum (FWHM)
with temperature is plotted for both dots. The lines are the
linear and parabolic fits for dot 1 and dot 2 respectively. Peak
width for dot 2 is multiplied by factor $10$ for visibility. The
peaks are close to $V_g=3$ V. b) Temperature dependence of the
peak-to-valley ratio for the fast oscillations. Solid (open)
symbols are for the ratios taken near the peaks (valleys) of the
slow oscillations.}
\label{\ffile}
\end{figure}

The sample is a Si quantum dot fabricated from a
silicon-on-insulator wafer.  A narrow bridge with a
lithographically defined dot is formed from the top Si layer; the
bridge is connected to wide source and drain regions via two
constrictions. Subsequently, a 50 nm thick layer of SiO$_2$ is
thermally grown around the dot, followed by a poly-Si gate. For a
detailed description of sample preparation see
Ref.~\cite{leobandung95a}.  In most cases, more than one dot is
naturally formed in such Si nanostructures, and the origin of
these additional dots is a subject of ongoing
research\cite{rokhinson00a,rokhinson00b,horiguchi01}. For this
particular study we have chosen a device that exhibits two
distinct periods of conductance oscillations at low temperatures.
As will be shown below, the device consist of two dots, both
participating in the charge transport.

The conductance $G$ through the sample is plotted in
Fig.~\ref{cond}a as a function of gate voltage $0<V_g<9$ V. The
temperature dependence is shown in the inset. At high
temperatures, $15$ K $<T<60$ K, the data is consistent with the
conventional theory of Coulomb blockade (CB) in a single
dot\cite{dotbook}. This dot will be called dot 1 throughout the
paper. At $T<15$ K the behavior of the conductance changes
qualitatively -- oscillations with a much smaller period  are
superimposed on the main dot CB oscillations. A remarkably large
number of these fast oscillations -- more than $500$ -- can be
resolved in a single scan. This coexistence of two periods
suggests that two dots are involved in the transport. In fact, we
can rule out interference effects as an origin of the fast
oscillations by showing that there is an electrostatic coupling
between the dots.

It is well known that quantum dots can be used as sensitive
electrometers, and, in our case, each of the dots can be
potentially used to measure the charge on the other dot. In
Fig.~\ref{trans} a normalized peak position of fast oscillations,
$V^{'}_p(N)=[V^p_g(N)/\langle\Delta V_{g2}\rangle-N]$, is plotted
as a function of $V_g$, where $V_g^p(N)$ is the position of the
$N$-th peak and $\langle\Delta V_{g2}\rangle=14$ mV is the average
peak spacing. For periodic oscillations $V^{'}_p$ should be a
constant, independent of $N$ (and $V_g$). Vertical lines mark
positions of the CB peaks in dot 1. Slips of $V^{'}_p$ appear
every time an electron is added into dot 1. These slips should be
expected for capacitively coupled dots: each electron, added into
dot 1, increases the electrostatic potential of dot 2 by
$\Delta\phi=\frac{eC_c}{C_{\Sigma1}C_{\Sigma2}}$, resulting in a
$\Delta\phi\frac{C_{\Sigma2}}{C_{g2}}$ shift of the CB peaks.
Here, $C_c$, $C_{g2}$, $C_{\Sigma1}$ and $C_{\Sigma2}$ are the
cross capacitance between the dots, the gate capacitance of dot 2,
and the total capacitance of dots 1 and 2. The slips are extended
over a few periods of the fast oscillations, due to the finite
broadening of the CB peaks in dot 1. We also observed two large
slips extended over $\sim20$ periods (see inset). These probably
reflect charging of some other traps, which do not participate in
the transport. In principle, slips in peak positions can occur for
an interferrometer geometry as well, as a result of the $\pi$
phase shift accumulated each time an electron enters the dot. This
effect would cause a slip of half a period per an added electron,
which is much larger than experimentally observed. Therefore, the
existence of the slips, correlated with CB peaks from dot 1, is an
unambiguous experimental evidence for the presence of the second
dot.

How are the dots coupled? If  the dots are strongly coupled
electrostatically, $C_c\gg C_{\Sigma1},C_{\Sigma2}$, the dimer
will behave as a single dot and the conductance should demonstrate
single period oscillations. This is clearly not the case in our
sample. In the opposite regime, $C_c\ll C_{\Sigma1},C_{\Sigma2}$,
one can distinguish between the two possibilities: i) the dots are
connected in parallel, and ii) the dots are connected in series.
At low gate voltages $V_g<3$ V the fast oscillations are
suppressed in the valleys of the CB in dot 1 (see left inset of
Fig~\ref{cond}a), implying that the dots are connected in series.

In the regime of sequential tunneling through two weakly coupled
dot connected in series the total conductance $G_{\rm
seq}^{-1}\sim G_1^{-1}+G_2^{-1}$, where $G_{1,2}$ are the
conductances of each dot\cite{ruzin92}.  Sequential tunneling is
in qualitative disagreement with the data. In such regime
amplitude of the fast oscillations $\delta R$ of the total
resistance $R_{\rm seq}=G_{\rm seq}^{-1}$ should originate from
the CB in the extra dot, $\delta R\approx G_2^{-1}$. As shown in
Fig.~\ref{cond}b, $\delta R$ is strongly correlated with $R_1$ and
changes by two orders of magnitude within a few periods of the
fast oscillations (there are $\sim 8$ periods of the fast
oscillations per slow period). Such strong modulations of the peak
height with such small ($\sim 8$ periods) correlation length is
not expected for either weakly\cite{chang96} or strongly
coupled\cite{patel98a} dots.

As Fig~\ref{cond}a shows, the amplitude of the fast oscillations
$\delta G$ is correlated with the envelope of the total
conductance $G_{\rm env}$: the amplitude is larger at the peaks
and smaller at the valleys. Indeed, the ratio $\delta G/G_{\rm
env}$, plotted in Fig.~\ref{cond}c, is practically
$V_g$-independent up to 6 V and gradually decreases with further
increase of $V_g$. Moreover, this ratio, rather than $\delta G$ or
$\delta R$, is almost the same near the minima and the maxima of
the main dot CB oscillations. This observation hints that the
total conductance should be treated quantum mechanically as a
transmission problem. In this case, the total conductance is
proportional to the {\it product} of the transmission through each
dot, $G_{QM}\propto\Gamma_{\rm total}=\Gamma_{1}\cdot\Gamma_{2}$,
and the transport through the whole dimer is {\it coherent}.

In the second part of the paper we use our knowledge of the
transport through the dimer to analyze some intriguing features in
the temperature dependence of the total conductance, and to show
that the transport is dominated by inelastic co-tunneling.
Parameters of dot 1 can be easily extracted in the usual
way\cite{dotbook}. The obtained gate capacitance $\sim 1-2$aF is
consistent with the geometrical estimates for the lithographically
defined dot. The charging energy $E_{c1}$, and the mean level
spacing $\Delta_1$ are extracted from the statistics of the peak
spacings, see left panel of Fig.~\ref{histo}. The spacings form a
broad distribution, consistent with $\Delta_1\sim E_{c1}\approx
4$meV. The mean level broadening is $\hbar\Gamma_{1}\approx0.8$
meV, less than $\Delta_1$. The peak widths strongly fluctuate at
low temperatures $k_BT<\hbar\Gamma_1$, where the width is
determined by the coupling to the leads. As expected from the
Random Matrix Theory, both distributions are asymmetric.

In Fig.~\ref{scalings} and the right panel of Fig.~\ref{histo} we
present the results of the $T$-scaling of peak width and the
distributions of peak spacing and width for dot 2, performed
similarly to the analysis of the main dot. The standard CB
analysis does not work for this dot: i) the peak widths do not
scale linearly with temperature; ii) the peak shapes are not
Lorentzian, although their width, $\Gamma_2$, saturates at low
temperatures at 4 mV, which translates to $\approx 2$ meV $\gg k_B
T$; and iii) there is no appreciable fluctuations of the peak
width even at low temperatures, where the $T$-dependence of each
peak has already saturated. Nevertheless, one can analyze the
distribution of the peak spacing. The sharpness of the
distribution requires the mean level spacing to be much smaller
than the charging energy, $\Delta_2 \ll E_{c2}$ ($E_{c2}$ does not
fluctuate). Thus, the following set of inequalities is satisfied
$\Delta_2, k_B T \ll \hbar\Gamma_2 < E_{c2}$. This means that dot
2 is in the strong coupling regime, where the standard CB theory
is not applicable. The above features of the second dot can be
understood within the framework the co-tunneling theory in the
strong coupling regime\cite{rev}.

There are two contributions to the conductance of a strongly
coupled asymmetric dot. One is from elastic co-tunneling (the dot
always remains in its ground state) and the other involves
inelastic processes, which create particle-hole excitations in the
dot. Elastic co-tunneling dominates at low temperatures, resulting
in a strongly fluctuating $G$ and, correspondingly, peak
width\cite{aleiner96}. At higher temperatures the leading
mechanism of electron transport is inelastic co-tunneling and the
conductance shows regular non-fluctuating periodic modulations as
a function of $V_g$, and gradually evolves with temperature from
CB peaks into smooth oscillations\cite{furusaki95}.  The relevant
energy scale is $E_c r^2\cos^2(\pi {\cal N})$, where $r$ is the
smallest of the reflection coefficients at the barriers, and
${\cal N}=V_g/eC_{g2}$ measures the charge that minimizes the dot
electrostatic energy, regardless of charge quantization. The
conductance at the maxima and minima of the oscillations are
estimated to be $G_{\rm max}\sim ({e^2}/{\pi\hbar}) k_B T r^2/(E_c
r^2)$, and $G_{\rm min}\sim ({e^2}/{\pi\hbar}) (k_B T)^2r^2/(E_c
r^2)^2$, respectively. The ratio between the maxima and minima is
expected to have a linear $T$-dependence
\[ G_{\rm min}/G_{\rm max}\approx k_BT/(E_{c} r^2).\]
Indeed, as shown in Fig.~\ref{scalings}b, the ratio $G_{\rm
min}/G_{\rm max}$ for the fast oscillations has linear
$T$-dependence regardless of whether it is measured near the peaks
(solid symbols) or the valleys (open symbols) of dot 1 CB
oscillations.  The fast oscillations are observed up to $k_B
T=E_{c} r^2$, thus strong coupling effectively renormalizes the
charging energy.

The anomalous temperature dependence of the conductance, observed
in the sample, can be naturally understood for the coherent
transport through the dimer. Experimentally, the conductance near
the peaks of dot 1 increases with temperature (see inset in
Fig.~\ref{cond}a), contrary to the prediction of the CB theory
that the peak height should be temperature independent for $k_B
T<\hbar\Gamma_1$ and $\propto 1/T$ for $\hbar\Gamma_1<k_B T
<\Delta_1<E_{C1}$ \cite{beenakker91}. Because transport is
coherent, the total conductance is $G\approx G_1\cdot
G_2/({e^2}/{\pi\hbar})$, and the anomalous temperature dependence
is a result of the $T$-dependence of $G_2$.  As we have shown
above, $G_2$ increases with $T$ due to the inelastic co-tunneling
processes. Thus, the observed anomalous temperature dependence is
an additional argument in favor of both coherent transport through
the dimer and inelastic co-tunneling in dot 2.

In conclusion, we proposed a new method to identify coherent
transport using two quantum dots, which does not require large
phase coherence length in the host material. We studied the
electron transport through a Si double-dot structure, and
established that the electrons are transferred coherently through
the whole dimer. Due to strong coupling to the leads of one of the
dots, the transport is dominated by the inelastic co-tunneling
processes, altering the conventional temperature dependence of the
CB oscillations. In the future, this method can be used to measure
$l_{\phi}$ in the interconnect region by varying the distance
between the dots.

This work was supported by the ARO, ONR and DARPA.


\begin{references}

\bibitem[*]{guoaddr}
Present address: Department of Electrical Engineering and Computer
Science,
  University of Michigan, Ann Arbor, MI 48109.

\bibitem{yacoby95}
A. Yacoby, M. Heiblum, D. Mahalu, and H. Shtrikman, Physical
Review Letters
  {\bf 74},  4047  (1995).

\bibitem{qd-stat-review}
For a recent review, see Y. Alhassid, Rev. Mod. Phys. {\bf 72},
895 (2000).

\bibitem{leobandung95a}
E. Leobandung, L.~J. Guo, Y. Wang, and S.~Y. Chou,
J.~Vac.~Sci.~Technol. {\bf
  13},  2865  (1995).

\bibitem{rokhinson00a}
L.~P. Rokhinson, L.~J. Guo, S.~Y. Chou, and D.~C. Tsui, \apl {\bf
76},  1591
  (2000).

\bibitem{rokhinson00b}
L.~P. Rokhinson, L.~J. Guo, S.~Y. Chou, and D.~C. Tsui,
Superlattices and
  Microstructures {\bf 5/6},  413  (2000).

\bibitem{horiguchi01}
S. Horiguchi {\it et~al.}, Jap. J. of Appl. Phys., Part 2 {\bf
40},  L29
  (2001).

\bibitem{dotbook}
For a review on quantum dots, see {\em Mesoscopic Electron
Transport}, Vol.~345
  of {\em {NATO} {ASI} Series}, edited by L.~P.~Kouwenhoven, L.~L.~Sohn and G.
  Sch{\"o}n (Kluwer, London, 1997).

\bibitem{ruzin92}
I.~M. Ruzin, V. Chandrasekhar, E.~I. Levin, and L.~I. Glazman,
\prb {\bf 45},
  13469  (1992).

\bibitem{chang96}
A.~M. Chang {\it et~al.}, \prl {\bf 76},  1695  (1996).

\bibitem{patel98a}
S.~R. Patel {\it et~al.}, \prl {\bf 81},  5900  (1998).

\bibitem{rev}
For a recent review see I.L. Aleiner, P.W. Brouwer, and L.I.
Glazman,
  cond-mat/0103008.

\bibitem{aleiner96}
I.~L. Aleiner and L.~I. Glazman, \prl {\bf 77},  2057  (1996).

\bibitem{furusaki95}
A. Furusaki and K.~A. Matveev, \prb {\bf 52},  16676  (1995).

\bibitem{beenakker91}
C.~W.~J. Beenakker, \prb {\bf 44},  1646  (1991).

\end{references}

\end{document}